\documentclass[sigconf,screen]{acmart}
\setcopyright{acmlicensed}
\acmDOI{10.1145/3650212.3652119}
\acmYear{2024}
\copyrightyear{2024}
\acmSubmissionID{issta24main-p402-p}
\acmISBN{979-8-4007-0612-7/24/09}
\acmConference[ISSTA '24]{Proceedings of the 33rd ACM SIGSOFT International Symposium on Software Testing and Analysis}{September 16--20, 2024}{Vienna, Austria}
\acmBooktitle{Proceedings of the 33rd ACM SIGSOFT International Symposium on Software Testing and Analysis (ISSTA '24), September 16--20, 2024, Vienna, Austria}
\received{16-DEC-2023}
\received[accepted]{2024-03-02}
\usepackage{subfig,graphicx}
\usepackage{enumitem}
\usepackage{bbding}
\usepackage{listings}
\usepackage{multirow}
\usepackage{makecell}
\usepackage{xcolor}
\usepackage{listings}
\usepackage{pifont}

\usepackage{xspace}
\newcommand{\toolname}{\textit{CustomDLCoder}\xspace}  %JG: suggested
\newcommand{\firstmodule}{\textit{Model Parsing}\xspace}
\newcommand{\secondmodule}{\textit{Computing Unit Extraction}\xspace}
\newcommand{\thirdmodule}{\textit{Configuring Data Analysis}\xspace}
\newcommand{\forthmodule}{\textit{Dynamic Configuration}\xspace}

\makeatletter

\DeclareRobustCommand\bmvaOneDots{\futurelet\@let@token\bmv@onedotsaux}
\def\bmv@onedotsaux{\ifx\@let@token.,\else.,\null\fi\xspace}
\DeclareRobustCommand\bmvaOneDot{\futurelet\@let@token\bmv@onedotaux}
\def\bmv@onedotaux{\ifx\@let@token.\else.\null\fi\xspace}
\makeatother
\def\eg{\emph{e.g}\bmvaOneDots} 
\def\ie{\emph{i.e}\bmvaOneDots}

\usepackage{tcolorbox}
\def\BibTeX{{\rm B\kern-.05em{\sc i\kern-.025em b}\kern-.08em
    T\kern-.1667em\lower.7ex\hbox{E}\kern-.125emX}}

\definecolor{codegreen}{RGB}{2,112,10}
\definecolor{codegray}{rgb}{0.5,0.5,0.5}
\definecolor{codepurple}{rgb}{0.58,0,0.82}
\definecolor{backcolour}{rgb}{0.95,0.95,0.92}

\usepackage{array} 
 
\lstdefinestyle{mystyle}{
  language=C++,
  backgroundcolor=\color{backcolour},   commentstyle=\color{codegreen},
  keywordstyle=\color{magenta},
  numberstyle=\tiny\color{codegray},
  stringstyle=\color{codepurple},
  basicstyle=\ttfamily\footnotesize,
  breakatwhitespace=false,         
  breaklines=true,                 
  captionpos=b,                    
  keepspaces=true,                 
  numbersep=5pt,                  
  showspaces=false,                
  showstringspaces=false,
  showtabs=false,                  
  tabsize=2
}

\begin{document}
	
\title{Model-less Is the Best Model: Generating Pure Code Implementations to Replace On-Device DL Models}

\author{Mingyi Zhou}
\orcid{0000-0003-3514-0372}
\affiliation{%
  \institution{Monash University}
  \city{Clayton}
  \country{Australia}
}
\email{mingyi.zhou@monash.edu}

\author{Xiang Gao}
\orcid{0000-0001-9895-4600}
\affiliation{%
  \institution{Beihang University}
  \city{Beijing}
  \country{China}
}
\email{xiang_gao@buaa.edu.cn}

\author{Pei Liu}
\orcid{0000-0001-6008-7265}
\affiliation{%
  \institution{CSIRO's Data61}
  \city{Clayton}
  \country{Australia}
}
\email{Pei.Liu@data61.csiro.au}

\author{John Grundy}
\orcid{0000-0003-4928-7076}
\affiliation{%
  \institution{Monash University}
  \city{Clayton}
  \country{Australia}
}
\email{john.grundy@monash.edu}

\author{Chunyang Chen}
\orcid{0000-0003-2011-9618}
\affiliation{%
  \institution{TU Munich}
  \city{Heilbronn}
  \country{Germany}
}
\email{chun-yang.chen@tum.de}

\author{Xiao Chen}
\orcid{0000-0002-4508-5971}
\affiliation{%
  \institution{University of Newcastle}
  \city{Callaghan}
  \country{Australia}
}
\email{xiao.chen@newcastle.edu.au}

\author{Li Li}\authornote{Corresponding author.}
\orcid{0000-0003-2990-1614}
\affiliation{%
  \institution{Beihang University, Beijing}
  \city{Yunnan Key Laboratory of Software Engineering}
  \country{China}
}
\email{lilicoding@ieee.org}




        \begin{abstract}
%Machine learning (ML) techniques have gained popularity in software development. 
Recent studies show that on-device deployed deep learning (DL) models, such as those of Tensor Flow Lite (TFLite),
%in real-world applications 
can be easily extracted from real-world applications and devices by attackers to generate many kinds of adversarial and other attacks. % and are easy to be attacked. 
Although securing deployed on-device DL models has gained increasing attention, no existing methods can fully prevent these attacks. %for on-device DL models. 
% In this work, we analyze the challenge of securing deployed TFLite models in real-world applications, (\ie the files containing computational graphs and weights) that are directly exposed to attackers. 
Traditional software protection techniques have been widely explored. If on-device models can be implemented using pure code, such as C++, it will open the possibility of reusing existing robust software protection techniques.
% one solution is removing this model representation and implementing the model inference process using pure code, \eg C++ codes. 
However, due to the complexity of DL models, there is no automatic method that can translate  
% inference process of
DL models to pure code.
To fill this gap, we propose a novel method, \toolname, to automatically extract on-device DL model information and synthesize a customized executable program for a wide range of DL models. \toolname first parses the DL model, extracts its backend computing codes, configures the extracted codes, and then generates a customized program to implement and deploy the DL model without explicit model representation. The synthesized program hides model information for DL deployment environments since it does not need to retain explicit model representation, preventing many attacks on the DL model. In addition, it improves ML performance because the customized code removes model parsing and preprocessing steps and only retains the data computing process.
Our experimental results show that \toolname improves model security by disabling on-device model sniffing. Compared with the original on-device platform (\ie TFLite), our method can accelerate model inference by \textbf{21.8\% and 24.3\%} on x86-64 and ARM64 platforms, respectively. Most importantly, it can significantly reduce memory consumption by \textbf{68.8\% and 36.0\%} on x86-64 and ARM64 platforms, respectively. %Our code extracting and refactor scheme has the potential to be a standard step for the DL model deployment.
% Our prototype tool can be found in an anonymous repository\footnote{\href{https://github.com/AnonymousAuthor000/code155}{https://github.com/AnonymousAuthor000/code155}}. 

\end{abstract}

        % \keywords{Code Generation, AI safety, DL optimization, SE for AI}
 
	\maketitle
        \section{Introduction}

More and more mobile applications (Apps) and IoT devices are leveraging deep learning (DL) capabilities.
Deploying DL models on such devices has gained great popularity as it avoids transmitting data and provides rapid on-device processing. It also enables applications to access their DL model offline.
As the computing power of mobile and edge devices keeps increasing, it reduces the latency of model inference and enables the running of large on-device models.

However, as such DL models are directly hosted on devices, attackers can easily unpack the mobile Apps, identify DL models through keyword searching, and then extract key information from the DL models.
This accessible model key information thus makes it easy to launch attacks or steal the model's intellectual property~\cite{zhou2024investigating}. 
To protect on-device models, the most commonly used on-device DL model framework, TFLite, converts the general DL model such as TensorFlow and PyTorch models to TFLite models, which disables direct white-box attacks. 
This is done by disabling the gradient calculation of the on-device models, which is essential for conducting effective white-box attacks. Such models are called non-differential models (\ie non-debuggable models).  
Other on-device platforms such as TVM~\cite{chen2018tvm} have similar processes. TVM compiles a high-level DL model representation into low-level representations that can be applied to various hardware platforms. It also supports packing such representation into the API library.
The low-level information makes it hard for attackers to reverse engineer the on-device model.

However, these on-device platforms still suffer from significant security risks. Recent attack methods~\cite{huang2022smart,li2021deeppayload,chen2022learning} can parse model information in the on-device model (\eg \texttt{.tflite}) files or the compiled low-level representation (\eg from TVM), then reverse engineer them or search for a similar debuggable one. Recent defence approaches propose to obfuscate the information of on-device models~\cite{zhou2023modelobfuscator}. The information inside model files (\eg .tflite files) is protected using several obfuscation strategies. However, the obfuscated model representation is still directly exposed to threats. Even if the obfuscated information is secured, the sniffing methods~\cite{xu2019first} can still locate the on-device model and generate black-box attacks~\cite{papernot2017practical,guo2019simple,zhou2020dast}, which is similar to the attacks for cloud models. \textbf{Therefore, as the DL model is easily attacked, the current mainstream model deployment strategy that employs explicit model information is a serious risk for mobile Apps and devices. }
In addition, model obfuscation will introduce inevitable overheads (\eg up to 20\% in RAM consumption) as it requires parsing the obfuscated APIs during model inference and injecting extra layers to achieve high obfuscation performance.

\textbf{Parsing the model file or identifying the DL component is usually required for the initial phase of attacking a DL model. we aim to explore whether we can hide DL components in their on-device deployment environment without introducing overheads to model inference.}
One approach is to deploy the DL model as a pure code implementation (\eg C/C++ code), eliminating the need for an explicit model representation that can be easily located and parsed.
It is also more efficient than the common model deployment strategies that deploy the library and explicit model representation separately.
For instance, \textit{m2cgen}~\cite{m2cgen} implements different ML algorithms as pure code by translating some ML algorithms to their corresponding pure code implementations. However, this strategy cannot be extended to DL techniques which have diverse architectures and frequently evolving algorithms.
Another alternative involves creating pure code implementations for specific DL algorithms such as \textit{llama.cpp}~\cite{llama.cpp}. However, this approach necessitates substantial manual effort for each DL model. Automatically generating code for complex ML algorithms is a difficult research problem. \textbf{Therefore, it is still required to design an automatic method that can translate the model inference to pure code implementations for various DL algorithms.} 
% \xiang{the above limitations do not mention scalability}
% However, a significant challenge arises from the current approaches that we do not have an automatic solution with high scalability. 

% that employ a model file to store DL model information and a generic API library to support model inference using the DL model file. These model files and API libraries are generated by open-source tools (\eg TensorFlow and PyTorch) and are easily identified on devices (\eg Android devices) using keywords~\cite{xu2019first}. 

% Inspired by existing AI compilers such as TVM, we propose a novel solution for compiling DL models, \toolname, to extract the computing code unit from the DL library and refactor them in an executable program. 

\begin{figure}[t]
  \begin{center}
    \includegraphics[width=\linewidth]{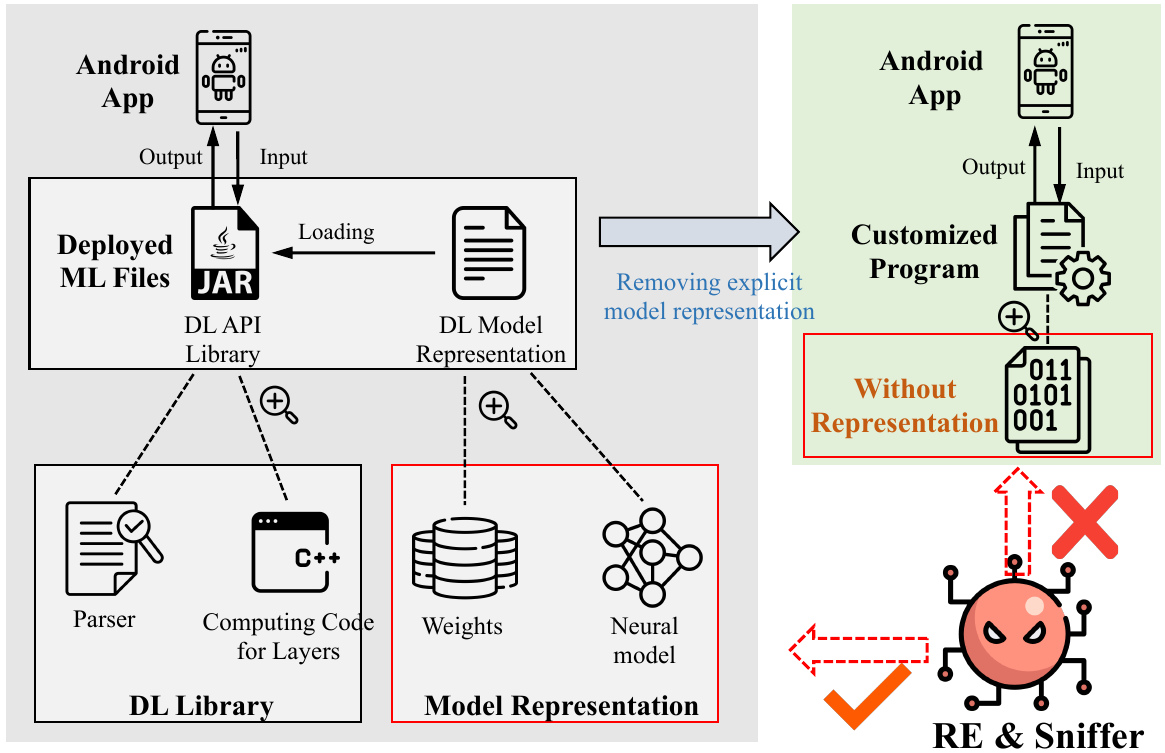}
  \end{center}
  \vspace{-1em}
  \caption{The high-level idea of generating pure code to replace DL model representations. The red block shows the difference between the deployed DL components.}
  \label{fig:overview}
\end{figure}

In this study, we want to answer the following research question: \textbf{Can we extract the essential codes from the DL API library and refactor them into an executable program for diverse DL algorithms?}
To this end, we propose a novel solution, \toolname, to extract the computing code unit from the DL library and refactor them in an executable program, whose overall architecture is shown in Figure~\ref{fig:overview}. 
Given a trained on-device DL model and its
underlying on-device DL library (TFLite), \toolname produces an executable C++ program that can be deployed on devices, as shown in the right part of Figure~\ref{fig:overview}.
\toolname removes the explicit model representations, \ie computational graph and weights, as illustrated in the left part of Figure~\ref{fig:overview}, hence avoiding the need of model representations for model inference. 
To achieve this goal, \toolname first parses the DL model, extracts its related computing codes from the TFLite library, configures the extracted codes, and then generates an executable program for model inference.  %In this paper, we explain the reason why current methods like static analysis are hard to solve this problem. Then, we propose four modules: \firstmodule, \secondmodule, \thirdmodule, and \forthmodule.
Our method will not affect the model's performance because it has the same computing process as the original model inference. Experiments on 11 representatives on-device DL models (\eg MobileNet~\cite{howard2017mobilenets}, SSD~\cite{liu2016ssd}, GPT2~\cite{radford2019language}) show that \toolname achieves a higher level of security compared to existing on-device protection methods without any overhead. In addition, the program generated by our method only contains the essential computing process for each model, by removing generic DL library steps involved in analyzing the computational graph and its parameters. This results in accelerating model inference (\textbf{by 21.8\% on x86-64 and 24.3\% on ARM64}) compared to existing deployment strategies, and reducing memory consumption (\textbf{by 68.8\% on x86-64 and 36.0\% on ARM64}).

The key contributions of this work include:
\begin{itemize}[leftmargin=*]
    % \item We propose to hide the DL-related components and features to achieve higher-level security than existing methods. It can prevent attackers from identifying the DL component on devices, and subsequently launching effective attacks.
    \item We propose a novel solution that can automatically extract the related backend code of specific models and refactor them to executable programs to address the challenge in existing methods;
    
    \item Our method can automatically translate various DL algorithms to pure codes, resulting in a higher level of security; 
    %. By preventing attackers from identifying the DL component on devices, we effectively thwart potential attacks. tools. %because the produce program just contains the computing-related process.
    %\item It can compile the TFLite model with only one command.
    \item Our methods can accelerate model inference (\textbf{by 21.8\% on x86-64 and 24.3\% on ARM64}) compared to existing deployment strategies and reduce memory consumption (\textbf{by 68.8\% on x86-64 and 36.0\% on ARM64}).
    \item We open-sourced our prototype tool \toolname by an GitHub repository: \href{https://github.com/zhoumingyi/CustomDLCoder}{https://github.com/zhoumingyi/CustomDLCoder}.
\end{itemize}

% We provide the source code of our prototype tool~\toolname  and our experimental datasets 

        \section{Background and Related Workds}

% Before introducing the details of our study, we first review essential background material and key terms used in our study.

%Our method analyzes the DL model and its corresponding DL library. The output of \toolname is the \textbf{customized DL program} which can be run without the model file. But it has the same outputs as the original model file under the same inputs. We use the \textbf{computing code unit} to define the computing code of each operator in TFLite, which only has the function of computing the output using inputs.

%\textbf{DL library} refers to the API library provided by the DL framework like TensorFlow and PyTorch. 

%The DL library contains the computing code of each layer. It needs to interact with the model file to compute the output using the given computational graph that is provided by the model file.
%DL library usually uses the \textbf{operator} to define the neural layer. One kind of neural layers may have multiple operators (\eg convolutional layer $rightarrow$ \texttt{conv2d, conv3d}).
%The \textbf{AI program} refers to the program generated by the AI compilers like TVM, that can be run on edge devices. The AI program just has essential computing APIs for model inference. It also needs to load the computational graph and weights.

%In this section, we will introduce the background of this study, which includes on-device DL frameworks, code obfuscation, as well as model obfuscation, 
% \cychen{Too much background, may shorten it so the audience can read the approach earlier.}

\subsection{DL Frameworks}

\paragraph{\textbf{Deep Learning (DL) Frameworks:}} The open-source community has developed many well-known \textbf{DL frameworks} to facilitate users to develop DL models, such as TensorFlow~\cite{tensorflow2015_whitepaper}, Keras~\cite{chollet2018keras}, and PyTorch~\cite{paszke2019pytorch}. These frameworks provide standards for developing DL models~\cite{dilhara2021understanding}. PyTorch is one of the latest DL frameworks which has gained academic user popularity for its easy-to-use and high performance. In contrast, TensorFlow is widely used by industry to develop new DL-based systems because it has the most commonly used on-device DL library, Tensor Flow Life (TFLite).  TFLite is the most popular library for DL models on smartphones, as it supports various hardware platforms and operation systems. 

\paragraph{\textbf{DL Deployment Strategy:}} As the training of DL models is intensive in both data and computing, mobile developers often collect data and train their models on the cloud or high-end desktop server prior to App deployment. 
Developers also need to  compile the trained models to be compatible with specific devices (\ie to produce on-device models) so as to speed up model inference on mobile CPU/GPUs~\cite{chen2021empirical,chen2020comprehensive,niu2021dnnfusion}. Developers use a tool (\eg \textit{TFLite Converter} in TensorFlow) integrated into their DL framework to compile their DL model to an on-device model. It will produce a compiled model that contains model architecture, weights and API library.
At App installation time, the compiled models and libraries are deployed, along with the App code itself, in the installation package of an App. At runtime, Apps perform the inference of DL models by invoking related APIs in their DL libraries.

\subsection{On-device DL Frameworks}

\paragraph{\textbf{On-device DL Frameworks:}} %TensorFlow provides many optimizations for compiling DL models to on-device models. 
TensorFlow provides a tool \textit{TensorFlow Lite Converter}\footnote{\href{https://www.tensorflow.org/lite/convert/index}{https://www.tensorflow.org/lite/convert/index}} to convert TensorFlow models into TFLite models. A compiled TFLite model can then be run on mobile and edge devices. However, it does not provide APIs to access the gradient or intermediate outputs like other DL models. 

%As illustrated in Figure~\ref{fig:overview}, 
Traditionally, on-device models are released as \textbf{DL files} that are deployed on devices.
Mobile app code then accesses these models through a dedicated \textbf{DL library}, such as the TFLite library if the AI model is developed using the TFLite framework.
Each model file contains two types of information: \textbf{computational graph} and \textbf{weights}, which record the model's architecture and parameters tuned based on the training dataset, respectively. 
Such a computational graph is usually a multi-layer neural network.
In the network, each layer contains an \textbf{operator} that accepts \textbf{inputs} (i.e., the outputs of the previous operator), \textbf{weights} (i.e., stored in the dedicated file that is pre-calculated in the training phase), and \textbf{parameters} (i.e., configuration of the operator. For example, the \texttt{conv2d} layer in TFLite requires the parameters of stride size and padding type. Their parameter will affect the outputs of layers.) to conduct the neural computation and outputs the results for the next operator.

As shown in Figure~\ref{fig:overview}, the traditional way of deploying DL models has to put DL model information directly on devices. The DL framework stores the model representations including computational graph and weights in one file (\eg \texttt{.tflite} file for TFLite) or packs them into the library when AI compilers such as TVM~\cite{chen2018tvm} are involved). However, these explicit model representations may be extracted and exploited by attackers~\cite{huang2022smart,li2021deeppayload,chen2022learning}, resulting in security threats to device users.

TFLite models run on the FlatBuffers Platform\footnote{\href{https://google.github.io/flatbuffers/}{https://google.github.io/flatbuffers/}}, which is efficient for loading the model and running it using multiple programming languages. It can access serialized data without parsing/unpacking and only needs small computational resources. TFLite uses the \texttt{.schema} file to define the data structures. For parsing the model structure and weights from the \texttt{.tflite} file, users can use the \texttt{.schema} file\footnote{\href{https://github.com/tensorflow/tensorflow/blob/master/tensorflow/lite/schema/schema.fbs}{schema file (The link is too long to display)}} of TFLite to parse the information on FlatBuffers level and get the JSON file that has detailed information of the \texttt{.tflite} model file. 

%\subsection{Deploying Deep Learning Models}

% \paragraph{\textbf{AI Compilers:}} AI compilers translate each atomic DL operator (\eg layers) into low-level representations and generate executable code by optimizing the computation and memory access for the target hardware devices. For more details about existing DL compilers, we refer readers to the existing studies~\cite{chen2018tvm, rotem2018glow, li2020deep, cyphers2018intel}. In this paper, we consider TVM~\cite{chen2018tvm}, a widely-used DL compiler to deploy DL models. TVM can compile the DL model into three components: computational graph (\ie model architecture), parameter (\ie weights), and shared library (\ie API library)\footnote{\href{https://tvm.apache.org/docs/arch/model_library_format.html}{https://tvm.apache.org/docs/arch/model\_library\_format.html}}. 

Compared with TFLite, TVM uses low-level representations (such as assembly code) to build the model, and it can then pack the representation into the library. However, those low-level representations still can be parsed by attackers~\cite{chen2022learning}. In addition, the performance of AI compilers like TVM relies on conversion accuracy. However, the conversion from high-level representation to low-level representation usually uses human-defined rules, which are often not stable due to the rapid change in DL frameworks. For instance, developers may have results inconsistency and conversion failure problems in TVM.

\subsection{DL Model Attacks}

DL models deployed on devices are subject to a range of attacks~\cite{papernot2017practical,zhou2020dast,li2021deeppayload,zhang2022investigating,huang2022smart,wu2024concealing}. These can include tricking the DL model with perturbed inputs into e.g. classifying an image incorrectly; extracting model information to facilitate other attacks; stealing a copy of the model (which may have been very expensive to produce) for use in one's own application; and others. These attacks can be black-box~\cite{wu2020decision,li2021deeppayload,huang2022smart} or white-box~\cite{zhang2022investigating}. Access to DL components and/or access to DL models facilitates these attacks.

\subsection{Code and Model Obfuscation}

\paragraph{\textbf{Code Obfuscation:}}
Code obfuscation methods were initially developed for hiding the functionality of malware. The software industry also uses it against reverse engineering attacks to protect code IP~\citep{schrittwieser2016protecting}. Code obfuscators provide complex obfuscating algorithms for programs like JAVA code~\citep{collberg1997taxonomy,collberg1998manufacturing}, including robust methods for high-level languages~\citep{wang2001security} and machine code level~\citep{wroblewski2002general} obfuscation. Code obfuscation is a well-developed technique to secure the source code. However, traditional code obfuscation approaches are hard to use to protect on-device models, especially for protecting the structure of DL models and their parameters. 

\paragraph{\textbf{Model Obfuscation:}}
To prevent attackers from obtaining detailed information of deployed DL models, model obfuscation has been proposed. This obfuscates key model information such as the semantic information of model components and model architectures~\cite{zhou2023modelobfuscator}. It is then hard for attackers to obtain the precise trained weights and structure of models from the obfuscated model because the connection between the obfuscated information and the original one is randomly generated. However, this method will introduce inevitable overhead to the model inference. Besides, this method cannot defend against black-box attacks for ML models, which are also effective in attacking the models, as they do not need the inner information of models. 
% We cannot prevent such attacks when attackers can locate the DL components in smart Apps or devices. 

\subsection{Customized DL Programs}

% We have two options to reduce the exposure of on-device DL components. First, we can fix the computational graph and weights into DL code to avoid attackers using model file keywords to identify the DL component. Second, we can use compilation techniques to remove the semantic information of the DL model, which can alleviate the information leakage by semantic extraction.

Our core idea is to replace DL model representation and generic libraries with {pure code} implementations (e.g., C++ Code, as shown in Figure~\ref{fig:overview}).
This code implementation, we refer to as a \textbf{customized DL program}. \textbf{Computing code units} define core inference processes for a DL operator. Note that one operator may have multiple computing code units.

We aim to hide the on-device DL representation of an app or device %to protect the AI system from  attacks %We combine the advantage of existing AI compilers and other on-device DL frameworks like TFLite to propose a new solution for
by extracting a DL model computing process into a customized DL program.
We generate customized C++ code implementations for the DL model inference that does not have explicit model information (\ie model, graph, and parameter files), and the model file or explicit model representation is unneeded in deployment. Because the generated code is a pure C++ program, we can use code obfuscation techniques to obfuscate the compiled program to only retain the computing process of the computational graph. %This will minimize the exposure of DL components on devices.  
%, can be obfuscated, and has the added advantage that it is more efficient than existing deployment strategies.
 %The pure code customized DL program can be obfuscated using traditional code obfuscation techniques to hide all the emboddied model structure.
This new customized, obfuscated DL program achieves the same objectives as of traditional DL representations + DL library approach.
However, it does not need the DL library anymore, and no model files or other representations need to be stored on insecure devices.
So, the DL functions are retained but deployed in a more secure manner.
In addition, as our generated code keeps the essential computing process and removes other model parsing steps, it can accelerate model inference and reduce memory consumption. 
% Compared to existing AI compilers like TVM, the customized DL model program generated will not need to analyze the computational graph and parameters, thus it can reduce the memory consumption for model inference.

%The approach can also improve the inference efficiency of the AI model, as lots of intermediate calls become unnecessary.
	
        % \input{ICSE2024/3_Preliminary}
        \section{\toolname Approach}
%\textbf{Difficulty} 
We want to generate pure code implementations of DL inference to replace on-device DL models and libraries. This should both improve their security to attacks and significantly improve their run-time efficiency. Although we can locate the code file for each DL operator easily, this study still has two main technical challenges. (1) First, it is hard to generate customized code implementations for DL models without efficiency loss. This is because the DL library contains various APIs for the same operator and numerous data preprocessing APIs to optimize the inference on different data types and devices. It is hard to exactly extract the optimal and minimized codes for each DL operator. The redundant extracted codes and redundant execution path will cause efficiency loss after converting the original DL model to the customized program. (2) Second, for removing redundant code, we only extract small pieces of code for each operator. That means we will extract a large number of small code fragments for large DL models (\eg large language model). Besides, for the sake of code performance, the extracted code is highly condensed and often incomplete. It is difficult to configure and assemble complex code pieces into an executable program using automated methods.

\begin{figure*}[t]
  \begin{center}
    \includegraphics[width=0.99\linewidth]{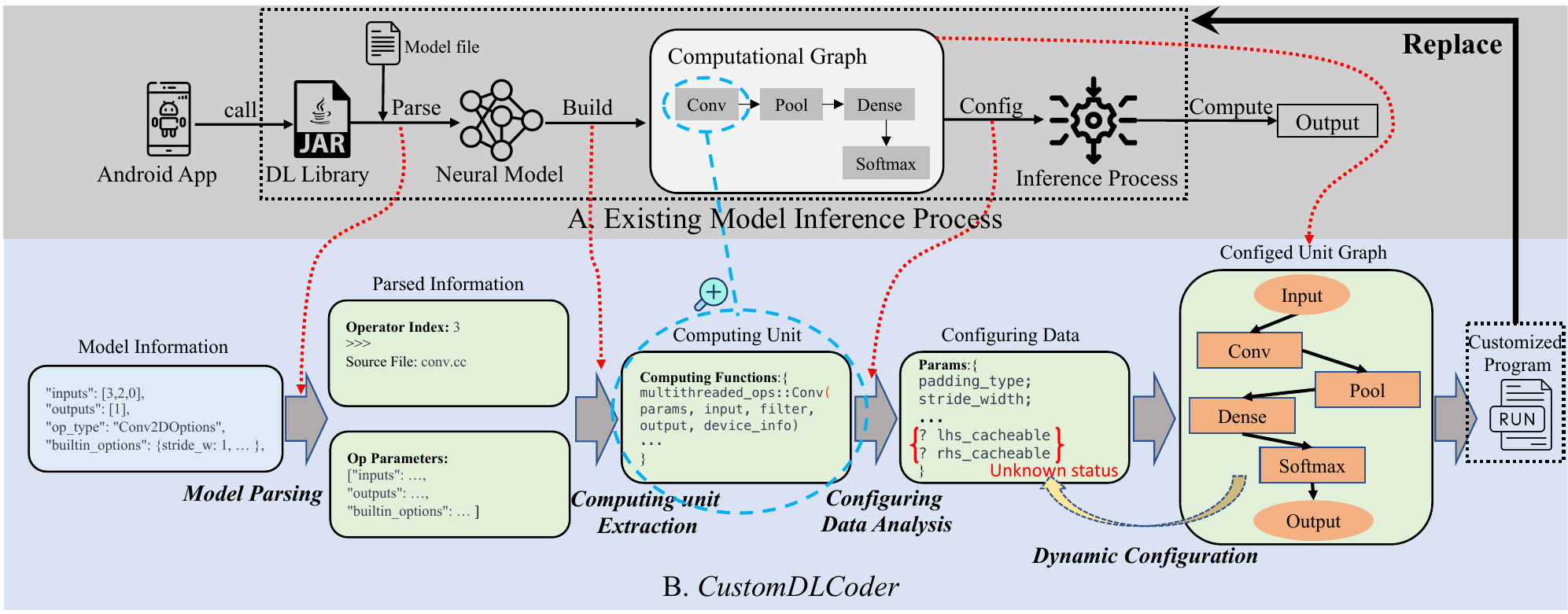}
  \end{center}
  \vspace{-1em}
  \caption{Design Overview of \toolname. The generated program is collected and confined by analyzing the inference process of the original TFLite library. }
  \label{fig:design_overview}
\end{figure*}

To solve the aforementioned problems, we argue we can trace the execution process of DL models in its related API library, extract the related small code units, and refactor them according to their original execution path and data structure. In addition, we propose a dynamic configuration algorithm to configure and assemble the incomplete small code pieces automatically. The overview of our proposed method is shown in Figure~\ref{fig:design_overview}. It has four main steps: (1) Model Parsing, (2) Computing Unit Extraction, (3) Configuring Data Analysis, and (4) Dynamic Configuration. We will detail our proposed method in the following subsections.

\subsection{\toolname Overview}
Given a DL model $\mathcal{G}(x)$ and the corresponding DL library, our goal is to extract the complete computing process of the library using the DL model information (\ie computational graph and weights) and compile it to a customized DL program $\mathcal{C}(x)$. Formally the procedure can be denoted as:
\begin{equation}
    \mathcal{C}(x) = \mathcal{G}(x) \qquad \qquad \quad \forall x\in \mathcal{X},
\end{equation}
For any inputs $x$ within the input range $\mathcal{X}$ of the task, the compiled executable program will output the same result as the original DL model using its target DL library. Moreover, the extracted program contains the same computing process of $\mathcal{G}$ (\ie the data flow from the input to the prediction) to maintain the model accuracy but simplify the model parsing and data processing. It thus will run faster than the original model inference using the DL library.

The key objectives of \toolname are to automatically identify the backend computing unit related to the operator in the given computational graph, refactor the extracted code, and then compile the refactored code to an executable program. As the generated code is a pure C++-based program, it can be further obfuscated by existing code obfuscation tools to remove any remaining semantic representations of the computational graph (\eg function names). 

\toolname carries out the following four steps: 1) \firstmodule: This parses the information contained in a DL model. It will analyze two main components, operators and parameters, of the DL computational graph. 2) \secondmodule: This will use the parsed information to identify the computing code unit in the DL library for each operator. These computing functions will be collected into a computing unit set. 3) \thirdmodule: After obtaining the computing unit set, we use the parsed parameter information of each operator to configure the data needed for each corresponding computing unit. 4) \forthmodule: Finally, this will refactor the collected computing functions and input data and assemble them into a complete C++ program according to the provided DL computational graph. For configurations that cannot be determined by given parameter information, which needs to be determined by the status of intermediate data or device, this module will further search for their configuration. Our proposed code extraction scheme will not affect the model performance and efficiency because the produced code program uses the same computing code to get the output.

\subsection{\firstmodule}
\label{sec:model_parsing}
The purpose of this step is to parse the information of DL models so that we can automatically extract the computing process of the given  DL model. The DL model file contains two kinds of information: operators and weights tensors. In this module, we first use \texttt{Flatbuffer} to reverse engineer the TFLite model file to a JSON file. This contains the high-level representation of the DL model, which is shown in the Parsed Information of Figure~\ref{fig:design_overview}. Note that the weights tensor of the model file will not be extracted in this step. After obtaining this high-level representation, we then use the operator index (that can be obtained from the operator information) to locate the source code for each operator. As shown in Listing 1, TFLite will register all operators in the `\texttt{register.cc}' file to assign the operator code (\eg \texttt{BuiltinOperator\_CONV\_2D = 3}) to TFLite operators or assign a name (\eg \texttt{Mfcc}) to custom operators as the operator index. Note that the custom operators are implemented by users. The computing code for each operator will be implemented in the \texttt{Register\_\{op\_name\}}. Thus, the TFLite interpreter can parse the operators in the computational graph by such operator index. We use the operator index extracted from the model file to collect the source code of operators that will be used in the model inference. For example, the operator code of \texttt{Conv2D} operator in TFLite is 3 (i.e., line 3). We then use this operator code 3 to identify the corresponding operator registration \texttt{Register\_CONV\_2D()}.
After that, we locate the source code, in the \texttt{conv.cc} file. For custom operators, we will use the operator name (\eg \texttt{Mfcc}) to identify the registration. For example, the custom operator with the name \texttt{Mfcc} will use \texttt{"Mfcc"} as the keyword to search the registration in the `\texttt{register.cc}' file.

\begin{lstlisting}[caption={The registration of operators in TFLite. The \texttt{Conv2D} is TFLite operator and the \texttt{Mfcc} is a custom operator implemented by users.},captionpos=b]
enum BuiltinOperator {
 // Operator code
 BuiltinOperator_CONV_2D = 3,
  ...
}

BuiltinOpResolver::BuiltinOpResolver(){
 // The registration for Conv2D operator
 AddBuiltin(BuiltinOperator_CONV_2D, Register_CONV_2D());
  ...
 // The registration for custom operator
 AddCustom("Mfcc", tflite::ops::custom::Register_MFCC());
}
\end{lstlisting}

\subsection{\secondmodule}

\begin{figure}[t]
  \begin{center}
    \includegraphics[width=0.8\linewidth]{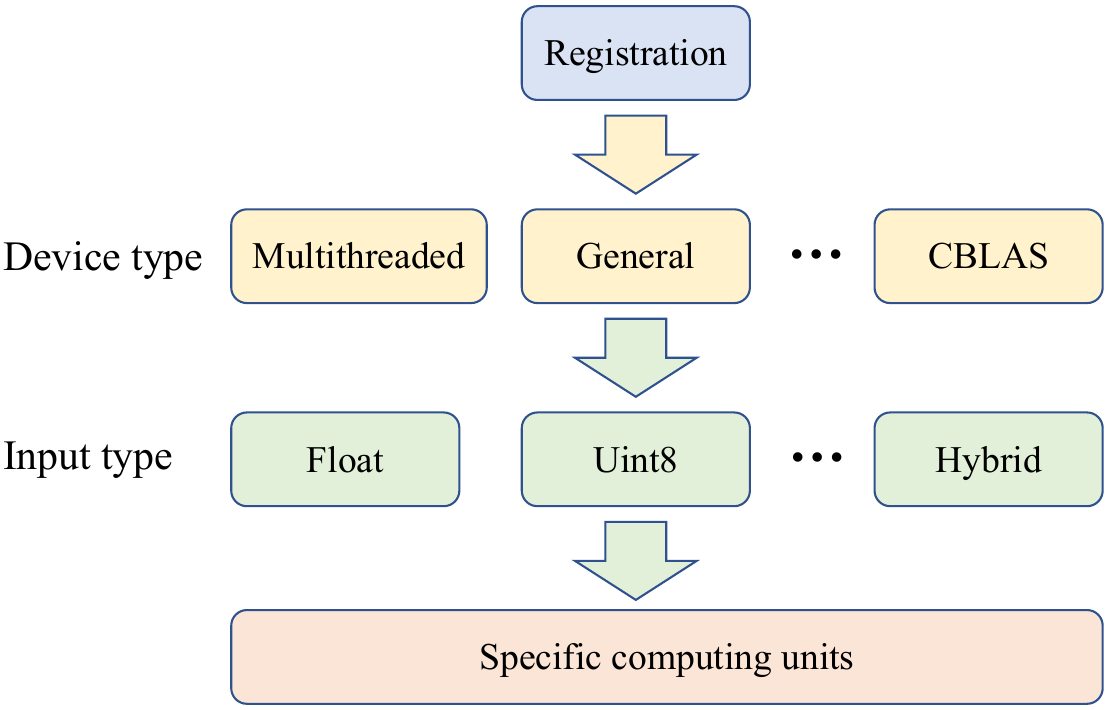}
  \end{center}
  \vspace{-0.7em}
  \caption{Structure of the operator source code. TFLite implements different code units for computing the output in different situations. It will parse the device and input information to choose the computing unit.}
  \label{fig:op_code_struc}
\end{figure}

After we collect the source code for each operator used in the DL model inference, we have all the backend computing code from the model input to the model output. However, the operator source code needs to support different hardware platforms and different data types (\eg \texttt{float}, \texttt{uint8}). For a given TFLite model, the model inference only uses part of the collected codes. We can remove the unneeded part of the collected code to improve the code efficiency. 

The function structure in an operator is shown in Figure~\ref{fig:op_code_struc}. The operator registration will first parse the device information to choose a specific registration entity. Each operator usually has several registration entities including a multithread-optimized entity, a general entity, and other entities that can be executed on special hardware platforms. Each entity has four functions: \texttt{Init}, \texttt{Free}, \texttt{Prepare}, and \texttt{Eval}. The TFLite interpreter will use \texttt{Init} and \texttt{Free} to initialize the data and delete the data in memory. When the TFLite interpreter loads the operator, \texttt{Prepare} function will be used to load and prepare the configurations (\eg output size) for this operator. \texttt{Eval} function computes the output using the given input, operator weights, and operator parameters. In the \texttt{Eval} function, it has different implementations for different data types. Thus, the source code of operators can be considered a set of computing code units that support different hardware platforms and different data types, which is shown in Figure~\ref{fig:op_code_struc}. Note that the computing code unit in TFLite has data manipulation in the matrix form, not the complex TFLite tensor form.

\begin{table}[t]
	\centering
        \small
	\begin{tabular}{lll}
		\toprule
		$\bf Algorithm~1 $ Computing Unit Extraction  \\
		\midrule

		\textbf{Input}:$ \text{ Device info } d, \text{ tensor info } t_n, \text{ computational graph } \mathcal{G},$ \\
		$ \text{ operator's source code } \textbf{U}_n = \{U_{n}^{i,j}\}, $ where $n$ is the id of operators.\\
		\textbf{Output}: Extracted computing unit collection $\overline{\textbf{U}}$.\\
            \textbf{Notation: } the n-th operator $O_n$, data type $dt$. \\
% 		$ \text{success rate for the attacks generated by D.} $ \\
            $1: \text{Initialize }  \overline{\textbf{U}}$ \\
		$2: \textbf{For } \ O_n \text{ in } \mathcal{G} \textbf{ do}:$ \\
		$3: \qquad \text{Parse the data type } dt  \text{ in } t_n  $ \\
% 		$3: \qquad \text{Update the substitute model}: $\\
            $4: \qquad \text{Extract the source code } \textbf{U}_n $ \\
            $5: \qquad \overline{u}_n = U_{n}^{dt,d}$ \\
		$6: \qquad \overline{\textbf{U}}\text{.append}(\overline{u}_n)  $ \\
		$7 :  \textbf{end For} $  \\
		$8: \textbf{Return } \overline{\textbf{U}}$  \\

% 		$7: \textbf{end for}  $ \\

		\bottomrule	
	\end{tabular}
	\label{tb:CUE}
	\vspace{-0.5em}
\end{table}

Therefore, we use computing unit extraction to obtain the computing process of each operator in the computational graph, as shown in Algorithm 1. First, we load the information of operators by parsing the computational graph and obtain the source code of each operator in the \firstmodule module. As we mentioned before, the source code of operators can be considered as a set of computing units that support different hardware devices and different data types. We then use the information about the target device and data type to identify the computing unit that will be used in the model inference. Finally, we iterate over all operators in the computational graph, and do the same extraction process above to get the computing unit set $\overline{\textbf{U}}$. However, the $\overline{\textbf{U}}$ is not a complete program. These computing units are like puzzle pieces, and we need to assemble these units into complete code implementations of model inference in subsequent modules. This is very hard to solve in normal programs. However, the computing code of DL libraries like TFLite only has a limited kind of configuration, which enable us to use rules to configure the code correctly.

\begin{lstlisting}[caption={Simplified computing unit of \texttt{Conv2d} operator when the data type is float and the hardware device supports the multithread-optimized implementation.},captionpos=b]
void EvalFloat {
  // Create essential data for the computing unit
  TfLiteTensor* input = create_input();
  TfLiteTensor* weights = create_weights();
  ConvParams params = create_params();
  // initialize the output tensor
  TfLiteTensor* output;
  // The called API for this computing unit
  multithreaded::Conv(params, input, weights, output);
}
\end{lstlisting}
\subsection{\thirdmodule}
% \paragraph{\textbf{\thirdmodule}}

Each computing unit has two kinds of data that need to be created to obtain its output: tensor (\eg the input and weights) and parameter. These data can be referred to as the configuration for the computing unit. As we extract the separate computing unit from each operator, we need to create such data to execute the DL model inference.  We use the \texttt{EvalFloat} unit in \texttt{conv2d} operator as an example, as illustrated in Listing 2. 

In TFLite, the tensor and parameter data are designed as \texttt{struct}s in C++. Our \toolname's  \firstmodule module will parse the operator information and tensor information. The weights tensor is easy to create, we just need to use the corresponding construction method to create it using the parsed tensor information. For the input tensor, we can get specifications of the input tensor from the parsed tensor information, and obtain the tensor value from the previous operator of the computational graph. We can then use them to construct the input data. For parameter data in TFLite (\eg \texttt{ConvParams params} as shown in Listing 2), this is produced by analyzing the operator information (\eg the \texttt{"builtin\_options"} in Figure~\ref{fig:design_overview}).

To create the parameter data automatically, we build a mapping function $f(p)$ for each operator, which is collected from the source code of TFLite. TFLite needs to use such functions to configure the parameters correctly in model inference. Creating the parameter data can then be formulated as $\texttt{params} = f_n(p)$, where $n$ is the operator index. We can use this method to create the most required data for each computing unit. 

However, as the examples shown in Figure~\ref{fig:design_overview} (\ie \texttt{lhs\_cacheable} of the configuring data), some members in parameter $\texttt{struct}$ data are determined by the device status and tensor status, which are not provided in the model representation. These members are referred to as unknown status. Therefore, we introduce a \forthmodule method to solve this problem.

% \begin{lstlisting}[caption={The definition of TFLite tensors.},captionpos=b]
% typedef struct TfLiteTensor {
%   // Quantization information. 
%   TfLiteQuantization quantization;
%   // A union of data pointers.
%   TfLitePtrUnion data;
%   // A pointer to a structure representing the
%   // dimensionality of data buffer.
%   TfLiteIntArray* dims;
%   // The data type specification for data.
%   TfLiteType type;
%   // How memory is mapped
%   TfLiteAllocationType allocation_type;
%   ...
%   // True if the tensor is a variable.
%   bool is_variable;
% } TfLiteTensor;
% \end{lstlisting}

\subsection{\forthmodule}

As the parameter $\texttt{struct}$ data has limited potential choices. A naive way to create configurations for unknown status is searching for an optimal setting in the whole space. However, this simple approach will usually be computationally infeasible when we search for an optimal configuration combination in the large DL model. This is because the large model may has too many operators that need to be configured. Therefore, we analyze the aforementioned status-related configuration and propose a \forthmodule method to automatically configure each computing unit in a computational graph. This module will produce the complete computing code for the given TFLite DL model. The process of \toolname's \forthmodule is shown in Algorithm 2. 

\begin{table}[t]
	\centering
        \small
	\begin{tabular}{lll}
		\toprule
		$\bf Algorithm~2 $ \forthmodule  \\
		\midrule

		\textbf{Input}:$ \text{ extracted computing unit } \overline{\textbf{U}} = \{\overline{u}_n\}, \text{ computational graph } \mathcal{G}.$  \\
		\textbf{Output}: a complete program $\mathcal{C}$\\
            \textbf{Notation: } the n-th operator $O_n$, the number of operators $s$ \\
% 		$ \text{success rate for the attacks generated by D.} $ \\
            $1: \text{Initialize a unknown status set } \textbf{M} $ \\
		$2: \textbf{For } \ O_n \text{ in } \mathcal{G} \textbf{ do}:$ \\
            $3: \qquad \text{find the unknown status } M_n  \text{ in } \overline{u}_n$ \\
		$4: \qquad \textbf{M}\text{.append} (M_n)  $ \\
% 		$3: \qquad \text{Update the substitute model}: $\\
		$5:  \textbf{end For} $  \\
            $6: \textbf{For } \ \{m_0, ..., m_s\} \text{ in } \{M_1, ..., M_s\} \textbf{ do}:  $ \\
            $7: \qquad \textbf{If } \text{check\_rules} (\{m_0, ..., m_s\})$: \\
            $8: \qquad \qquad \text{Configure the } \overline{\textbf{U}} \text{ using } \{m_0, ..., m_s\} \text{ to get } \mathcal{C}$\\
            $9: \qquad \qquad \textbf{If } \| \mathcal{C}(x) - \mathcal{G}(x) \|_2 < \delta$ \\
		$10: \qquad \qquad \qquad \textbf{Break } $  \\
            $11: \textbf{Return } \mathcal{C}$ \\

% 		$7: \textbf{end for}  $ \\

		\bottomrule	
	\end{tabular}
	\label{alg:dynamic_config}
	\vspace{-0.5em}
\end{table}

After we obtain the computing unit set from the \secondmodule and configure its known data using the \thirdmodule, we then collect any unknown status for all operators in the computational graph. After that, we use status check rules to find eligible configurations from all possibilities of configurations. The status check rules are designed to reduce the search space. According to our analysis of collected DL models, we define two status check rules: 1) the unknown status in two computing units should be the same when the two computing units have the same operator information (\ie name and known parameters). This is because, in DL models, the same kind of operators in one model usually have the same data type and configurations. 2) The same kind of data (\ie input, weights, \texttt{Conv2d} parameter) should have the same unknown status. This is because TFLite usually gets the same system and tensor status information for the same kind of data. The two rules can significantly reduce the search space because DL models usually are a combination of basic model blocks. For example, a well-known model architecture ResNet is built with a lot of similar residual blocks, which only have limited kinds of operators and data. If the configuration is eligible, we assemble the computing unit set to a complete code implementation for the model inference. We iterate over all eligible unknown configurations until the output difference between the generated program returns and the original model inference is lower than a prefixed value $\delta$ under the same input. The output difference can be calculated by:
\begin{equation}
    diff = \| \mathcal{C}(x) - \mathcal{G}(x) \|_2
\end{equation}
where $\mathcal{C}$ is produced computing program, $\mathcal{G}$ is the original model inference. $x$ is the input that is in the eligible range of this model. We use the $l_2$-norm to measure the distance between two outputs. In experiments, we use 100 random inputs to compute the output difference. When the generated program has a similar output to the original model inference, the generated program can be used to replace the model file and library in the deployment environment.

\subsection{Compilation}
Finally, the generated C++ program from \toolname embodies the complete computing process as the original model inference, without needing the DL model file or representations. Because this generated code is a pure C++ program, we can use existing code obfuscation techniques to obfuscate the executable program to remove the semantic information and just keep its computing process. The DL component in the compiled program is thus hard to be identified and its DL model information cannot be easily decompiled by attackers. Code obfuscation is a well-developed technique, we omit discussing it in this study.

        \section{Evaluation}

Our \toolname aims to provide better security and performance for on-device deployed DL solutions. To determine if this has been achieved, we evaluate \toolname by answering the following key research questions:

\begin{itemize}[leftmargin=*]

 \item \textbf{RQ1: } How accurate is \toolname in transforming on-device models to executable programs?

 \item \textbf{RQ2: } How effective is \toolname in defending against DL model information parsing and extraction?

 \item \textbf{RQ3: } How efficient is the program generated by \toolname at model inference compared with original TFLite?

% \item \textbf{RQ2: } How effective is our approach in defending against the model extraction?

\item \textbf{RQ4: } What is the memory cost of code generated by \toolname compared with original TFLite?

\item \textbf{RQ5: } Is the code produced by \toolname better than other model deployment strategies.
% \item \textbf{RQ4: } Which obfuscation strategy affects the size of produced DL library?

 \end{itemize}

 %In this section, we show the evaluations of the proposed \toolname in terms of correctness, latency, memory efficiency, and reliability. 
 
 % Our experimental settings are shown as follows:

 \paragraph{\textbf{Dataset}} To evaluate \toolname's performance on models with various structures for multiple tasks, we use the DL model collected in Kaggle Model Hub\footnote{\url{https://www.kaggle.com/models}} and Huggingface\footnote{\url{https://huggingface.co/models}} to evaluate our proposed method including the fruit recognition model, the skin cancer diagnosis model, MobileNet~\cite{howard2017mobilenets}, MNASNet~\cite{tan2019mnasnet}, SqueezeNet~\cite{iandola2016squeezenet}, EfficientNet~\cite{tan2019efficientnet}, MiDaS~\cite{ranftl2020towards}, Lenet~\cite{lecun1998gradient}, PoseNet~\cite{kendall2015posenet}, SSD~\cite{liu2016ssd}, and GPT-2~\cite{radford2019language}, respectively.  

  % \paragraph{\textbf{Baseline}} We use the TFLite library and its converted TVM program as the baselines.

\paragraph{\textbf{Experimental Environment}} %The purpose of this \toolname evaluation is to demonstrate the 
\toolname is evaluated on a workstation with Intel(R) Xeon(R) W-2175 2.50GHz CPU, 32GB RAM, and Ubuntu 20.04.1 operating system and a Xiaomi 11 Pro smartphone with Android 13 OS.

\paragraph{\textbf{Setting}} For comparing the latency between our method and others, we run the model for one sample to compute the time consumption because some DL platforms may have optimization methods for the input with multiple samples, \ie keep the essential intermediate data in memory. To accurately compute the time consumption, we repeat the process 1,000 times and compute the average latency on x86-64 platforms. For computing the memory usage, we use the \textit{Valgrind Massif}\footnote{\url{https://valgrind.org/docs/manual/ms-manual.html}}. \textit{Valgrind} is a powerful instrumentation framework for memory profiling. \textit{Massif}, a \textit{Valgrind} tool, can be used to measure memory usage accurately.
% For the original TFLite platform, we use the \textit{CMake} tool to compile the TFLite API library as the baseline because \toolname also uses \textit{CMake} to produce the executable program.

\begin{table*}[t!]
% \small
\small
% \vspace{-1.0em}
\centering
\caption{The maximal translation error $(\times 10^{-5})$ of our proposed \toolname and other model deployment strategies on \texttt{x86-64} platform. \toolname program, TVM model, and ONNX-Runtime model are all converted from the TFLite model.}
\begin{tabular}{c|lccccccccccc|c}
\hline
\multirow{4}{*}{error} & Model name    & Fruit & Skin cancer &MobileNet&MNASNet &SqueezeNet &EfficientNet &MiDaS &Lenet &PoseNet &SSD & GPT-2 & Average \\
\cline{2-14}
& TVM     & 0.40 & 0.13 & 0.21     &  1.03  & 0.17      & 0.07    &  73.24 &  0.05       & 4.20    & 1.86   & 34.33  & 10.52\\
& ONNX-Runtime     & 52.72 & 38.31 & 0.01      & 0.51   &  0.23    & 36.89    & 101.23  & 0.01        &  11.7   & 1.62   & $4.6\times10^5$  & $4.2\times10^4$ \\
% \hline
& \toolname  & \textbf{0.0} & \textbf{0.0} & \textbf{0.0}    & \textbf{0.0}   & \textbf{0.0}      & \textbf{0.0}    &  \textbf{0.0} &  \textbf{0.0}       &\textbf{0.0}  &   \textbf{0.0}   & \textbf{0.0}  & \textbf{0.0}\\
\hline
\end{tabular}
\label{tb:trans_acc}
\end{table*}

\begin{table*}[t]
% \vspace{-1.0em}
 \small
\centering
\setlength{\tabcolsep}{2pt}
%\caption{The success number of existing model conversion tools and attack extracting the information of models. `\textit{Basic obfuscation}': \textit{renaming} + \textit{parameter encapsulation}.}
\caption{The success number of existing attacking methods to parsing the information of on-device models. `$\surd$': this information collection method cannot extract information for all models. `-': not applicable for this method. We use `\XSolidBrush' to show that TVM can be attacked by reverse engineering the low-level representation of computation graphs with NNReverse~\cite{chen2022learning}. However, this method is only effective for the TVM. We omitted it from our experiments and simply used its results.} 
\vspace{-0.5em}
\begin{tabular}{l|cccc|c|c|c}
\toprule
                                       &  \multicolumn{4}{c|}{\bf Model conversion tool} & \multicolumn{1}{c|}{\bf Reverse Engineering}  & {\bf Feature analyzing} &  {\bf Searching}\\
\hline
                                       & TF-ONNX & TFLite2ONNX & TFLite2TF & ONNX2TF & Reverse Engineering~\cite{li2021deeppayload,chen2022learning} & Smart App Attack~\cite{huang2022smart} &  DL Sniffer~\cite{xu2019first} \\
\hline
Original TFLite (without defense)                      & 11              & 10          & 10       &   -       & 11           & 8    & 11     \\ 
% \hline
\hline
 Model Obfuscation\cite{zhou2023modelobfuscator}     & $\surd$               & $\surd$           & $\surd$    &     -        & $\surd$            & $\surd$       & 11    \\ 
TVM     & -              & -           & -          &  -     & \XSolidBrush            & $\surd$    & 11      \\ 
ONNX-Runtime & -               & -           & -        &   10      & 10            & 8    & 11      \\ 
\textbf{\textit{\toolname}}     & -               & -           & -       & -         & $\surd$            & $\surd$     & $\surd$     \\ 
% \midrule 
\hline
\end{tabular}
% \vspace{-1em}
\label{tb:defence}
\end{table*}

\paragraph{\textbf{Baseline}} As our proposed \toolname extracts the code from the TFLite platform, we use the TFLite platform as the baseline, \ie using \texttt{.tflite} model file and the API library compiled by \texttt{CMake}, because \toolname also uses \textit{CMake} to produce the executable program. The compilation process can be found on the TFLite official documents\footnote{\url{https://www.tensorflow.org/lite/guide/build_cmake}}. In addition, we also use other model deployment strategies (\ie TVM, ONNX-Runtime) to show the efficiency of our method in compiling the DL models.

\subsection{RQ1: Transformation Accuracy}
We need to first evaluate the compilation correctness of our generated customized DL code. Table~\ref{tb:trans_acc} summarises the evaluation of our methods on compilation correctness. If the compilation has errors (\ie the conversion is not correct), the performance of DL models will lose. The maximal compilation error can be formulated as:
\begin{equation}
    \theta = \max _{i=1}^N\|\mathcal{C}\left(x_i\right)-\mathcal{G}\left(x_i\right)\|
\end{equation}
where $\mathcal{G}$ is the model inference before compilation. $\mathcal{C}$ refers to compiled program. We use 100 inputs to compute the maximal compilation error. 
Our \toolname method does not have any compilation errors as \toolname uses the same computing process as the original model inference.

In comparison, TVM and ONNX-Runtime will have slight compilation errors because they rely on conversion rules to compile a high-level representation to a low-level representation, which will cause inevitable compilation errors. As per our observation, for the model with a more complex architecture, TVM and ONNX-Runtime models will have more compilation errors, as the error for each operator will be accumulated in the next computing.

\begin{tcolorbox}[colback=gray!5!white,colframe=gray!85]
RQ1 Answer: \toolname will not introduce any compilation errors to the generated program and has better compilation correctness than other deployment strategies. 
\end{tcolorbox}

\subsection{RQ2: Resilience to Attacks}

\begin{table*}[t]
\small
% \footnotesize
% \vspace{-1.0em}
\centering
\caption{The latency (ms per input) compared with the original TFLite library. \toolname programs are converted from the TFLite model. The lower is better. `x86-64': the experiments on the Ubuntu workstation. `ARM64': the experiments on the Xiaomi 11 Pro smartphone with Android 13.}
\begin{tabular}{c|lccccccccccc|c}
\hline
\multirow{3}{*}{x86-64} & Model name    & Fruit & Skin cancer &MobileNet&MNASNet &SqueezeNet &EfficientNet &MiDaS &Lenet &PoseNet &SSD & GPT-2 & Average \\
\cline{2-14}
% & ONNX-Runtime & & & & & & & & & & & & \\
% & TVM     & \textbf{30} & 92 & 55     &  105  & 129      & \textbf{92}    &  647 &  26       & 113    & 217 & & 150   \\
& TFLite (Baseline)     & 33.0 & 98.6 & 60.4     &  83.8  & 58.9    & 99.1    &  484.8 &  2.9 & 116.6    &227.9 & 578.8 & 167.7    \\
& \textbf{\toolname}  & \textbf{32.6} & \textbf{91.9} & \textbf{54.6}     & \textbf{70.8}   & \textbf{50.2}      & \textbf{84.9}    &  \textbf{317.0} &  \textbf{2.0}       & \textbf{110.3}    &   \textbf{195.9}  & \textbf{433.0} & \textbf{131.2}   \\
% (0, 20)  & - & - & -     & -   & -      & -    &  - &  -       & -    & -  & \textbf{-}        \\
\hline
\multirow{2}{*}{ARM64} & TFLite (Baseline)    & 13.1 & 42.9  & 26.9     &  34.8  &  52.9   &  40.0   & 406.8  & 5.1  &  \textbf{42.7}   & 96.1 & 530.2  & 117.4    \\
& \textbf{\toolname}  & \textbf{12.7} & \textbf{35.1} &   \textbf{22.8}  &  \textbf{27.8}  &  \textbf{52.4}    & \textbf{33.8}      & \textbf{272.1}  &  \textbf{1.4}      & 44.0    &   \textbf{84.9}  & \textbf{391.6} & \textbf{88.9}  \\
\hline
\end{tabular}
\label{tb:time}
\end{table*}

We use the potential attacks mentioned in~\cite{zhou2023modelobfuscator} to evaluate the effectiveness of \toolname in concealing information in DL models. In addition, we use the \textit{DL sniffer} to show the performance of deployment strategies in defending against keyword searching~\cite{xu2019first}. We use methods to collect model information as follows:

\begin{enumerate}[leftmargin=*]
    \item \textbf{Model format conversion:} Existing model conversion tools first parse the model's structure and weights, and then assemble them into a new model with different formats. We utilize four tools, namely TF-ONNX~\citep{tf2onnx}, TFLite2ONNX~\citep{tflite2onnx}, TFLite2TF~\citep{tflite2tensorflow}, and ONNX2TF~\citep{ONNX2TF} to evaluate the performance of different methods. If a tool can convert the model to the target model format, we consider it a successful model information extraction. Note that the results may change for different versions of tools. Conversely, if the defence method is effective, these tools will be unable to extract the model information.
    \item \textbf{Reverse engineering for in-model information:} A TFLite model can be parsed by the \texttt{Flatbuffer}. The other model representation types can also be collected. We refer to a study in ~\citep{li2021deeppayload,chen2022learning}, in which the researchers attempt to extract a model's structure and weights inside the model file. 
    \item \textbf{Finding a similar model from the Internet:} This involves comparing the features among models. Attackers can use the App Attack to find a debuggable model with a similar structure and weights from the Internet~\citep{huang2022smart}. For the App Attacking method, if it can correctly identify the model with defences that have the same model structure as the original one on TensorFlow Hub (8 models in our test set are collected from the TensorFlow Hub), we consider it a successful model extraction.
    \item \textbf{Ientifying DL model and model components using keyword searching: } This includes searching for operators (\eg `\texttt{conv2d}') and weights in model representations (\eg `\texttt{.tflite}' file). We adopt the method in \cite{xu2019first} to try to identify the model representation in App packages. In our test set, two models were originally collected from the Android Apps. For other models, we randomly pack them into the public Andoird Apps as the test set. In experiments, we use \textit{Apktool} to unpack App packages and use the DL sniffer to search the DL-related component.
\end{enumerate}

Our attack results are shown in Table~\ref{tb:defence}. Model obfuscation was recently developed to obfuscate on-device DL model information. It can successfully defend against all attacking methods except for the searching technique. This is because it will use the conventional model-library interaction for model inference. Such interaction needs a model file (\ie `\texttt{.tflite}') to store the computational graph. Thus, attackers can use the keyword of model format to search for the corresponding model file. However, our \toolname can compile the DL model to an executable program, which does not use any semantic representations to store the model information. Therefore, the parsing and reverse engineering tool cannot be applied to the proposed deployment method. 
In addition, our method can use traditional existing code obfuscation methods to enhance the model security because the program generated by our tool is the complete C++ code.
In comparison, other AI compiler methods like TVM cannot achieve this. TVM needs files to store its model representation (\ie computational graph) and weights using low-level representations or such information can be extracted from the program because the TVM program needs to interact with the representation of the computation graph and weights. Although TVM uses low-level representations, such low-level representations can be converted to the corresponding high-level representation~\cite{chen2022learning}. It cannot resist such reverse engineering tools and search by keywords. The ONNX-Runtime uses the ONNX model that can be converted from TFLite models. However, ONNX models can be also converted to the TFLite model. So, the ONNX-Runtime platform can be attacked by the methods that are effective for TFLite models.

\begin{tcolorbox}[colback=gray!5!white,colframe=gray!85]
RQ2 Answer: Programs produced by \toolname can mitigate attacking methods that use semantic information to identify DL components. It provides the deployed DL model a higher level of security compared with existing deployment methods by encoding model file information in generated code and obfuscating this generated code.
\end{tcolorbox}

\begin{table*}[t]
% \small
\small
% \vspace{-1.0em}
\centering
\caption{The size (Mb) of deployed components for different deployment methods on the \texttt{ARM64} platform. TFLite components contain the model file and the compiled API library (be \textit{CMake}). Our \toolname only has an executable file that also compiled by \textit{CMake}.}
\begin{tabular}{lccccccccccc|c}
\hline
 Model name    & Fruit & Skin cancer &MobileNet&MNASNet &SqueezeNet &EfficientNet &MiDaS &Lenet &PoseNet &SSD & GPT-2 & Average \\
\cline{2-13}
TFLite (Baseline)    & 58.6 & 70.0 & 63.4    &  70.6  & 58.1      & 71.7    &  119.4 &  59.6       & 25.6    & 58.1 & 378.4  & 94.0   \\
% & TVM     & 27 & 52 & 30     &  43  & 36     & 53    &  \textbf{126} &  21       & 35    & 62 & &  49  \\
 \textbf{\toolname}   & \textbf{15.0}  & \textbf{26.4} & \textbf{19.3}     & \textbf{25.9}   & \textbf{14.2}      & \textbf{28.3}    &  \textbf{74.4} &  \textbf{14.5}       & \textbf{12.2}    &   \textbf{35.2} & \textbf{308.9} & \textbf{52.2}    \\
% (0, 20)  & - & - & -     & -   & -      & -    &  - &  -       & -    & -  & \textbf{-}        \\
\hline
\end{tabular}
\label{tb:size}
\end{table*}

\subsection{RQ3: Efficiency of Generated ML Code}

Our proposed \toolname extracts the computing code from the TFLite library. It reduces the computational complexity because \toolname removes some inference-unrelated steps (\eg analyzing the computational graph and operator parameters) during code generation. However, TFLite models use \texttt{Flatbuffer} to save and load the model. Parsing and loading the file using \texttt{Flatbuffer} is very fast and can increase the efficiency of the model inference. The source code of TFLite is optimized for running the model on \texttt{Flatbuffer}. We need to experiment to show the inference time efficiency of our generated DL model code is at least as good.

\begin{table*}[t]
% \small
\small
% \vspace{-1.0em}
\centering
\caption{The RAM consumption (Mb) of model inference for different deployment methods. \toolname programs are converted from the TFLite model. The lower is better. Note that we use the peak memory cost to show the results.}
\begin{tabular}{c|lccccccccccc|c}
\hline
\multirow{4}{*}{x86-64} & Model name    & Fruit & Skin cancer &MobileNet&MNASNet &SqueezeNet &EfficientNet &MiDaS &Lenet &PoseNet &SSD & GPT-2 & Average \\
\cline{2-14}
& TFLite (Baseline)    & 20.41 & 45.84 & 31.55    &  39.80  & 29.26      & 43.97    &  220.22 &  12.40       & 25.60    & 72.33 & 418.33  & 87.25   \\
% & TVM     & 27 & 52 & 30     &  43  & 36     & 53    &  \textbf{126} &  21       & 35    & 62 & &  49  \\
& \textbf{\toolname}   & \textbf{8.89} & \textbf{11.95} & \textbf{10.47}     & \textbf{9.65}   & \textbf{17.74}      & \textbf{11.95}    &  \textbf{94.30} &  \textbf{5.96}       & \textbf{14.73}    &   \textbf{19.01} & \textbf{94.99} & \textbf{27.24}    \\
% (0, 20)  & - & - & -     & -   & -      & -    &  - &  -       & -    & -  & \textbf{-}        \\
\hline
\multirow{2}{*}{ARM64} & TFLite (Baseline)    & 15.35 & 32.89 & 21.47     & 29.74   & 27.55    & 32.07    &  219.61 & 13.29   & 20.16    & 48.37 & 483.96 & 85.86    \\
& \textbf{\toolname}  & \textbf{11.98} & \textbf{24.45} & \textbf{17.23}    & \textbf{23.30}   &  \textbf{13.94}    & \textbf{25.78}    & \textbf{99.18}  &  \textbf{11.32}      &  \textbf{13.64}   &  \textbf{40.94}   & \textbf{322.85}& \textbf{54.96}  \\
\hline
\end{tabular}
\label{tb:mem}
\end{table*}

To show the performance of \toolname generated DL code, we compare our method with the original TFLite platform. The results are shown in Table~\ref{tb:time}.
% To show the performance of \toolname generated DL code, we compare our method with the existing AI compiler TVM, which is a well-known tool to compile DL models to low-level representations and programs. The results are shown in Table~\ref{tb:time}.
Note that we do not use any additional optimization for each deployment method, and we use multi-thread mode on the \texttt{x84-64} platform. The model runs faster on ARM64 because TFLite optimizes the model performance on ARM64, and we run the model using Python on \texttt{x86-64}. In contrast, we build executable programs using C++ to generate samples to compute the inference time.
Our method can accelerate model inference by 21.8\% and 24.3\% on x86-64 and ARM64 platforms, respectively.
Our \toolname generated code has the lowest average inference time because it is much more efficient on the largest neural networks (\ie MiDaS and GPT-2). TFLite achieves the fastest model inference in the tiny model PoseNet. This is because loading and configuring the model information costs most of the time in model inference for some operators, and the \texttt{Flatbuffer} is efficient in doing it. \textbf{In most cases, our proposed \toolname achieves better time efficiency than the original TFLite platform because it only contains essential output computing code to execute the model inference}.

\begin{tcolorbox}[colback=gray!5!white,colframe=gray!85]
RQ3 Answer: Our \toolname can accelerate the model inference compared with the original TFLite model. In addition, the program generated by our approach runs much faster than TFLite models for large neural networks.
\end{tcolorbox}

% \begin{table}[t]
% % \small
% \small
% % \vspace{-1.0em}
% \centering
% \caption{The size of deployment files (including model file, library, program). The lower is better.}
% \begin{tabular}{lcccccccccc|c}
% \toprule
%  ID    &\ding{192}&\ding{193}&\ding{194}&\ding{195} &\ding{196} &\ding{197} &\ding{198} &\ding{199} &\ding{200} &\ding{201} & Avg \\
% \midrule
% TFLite     & 19 & 49 & 33     &  49  & 41      & 54    &  238 &  \textbf{7}       & 42    & 98  & 63   \\
% TVM     & 27 & 52 & 30     &  43  & 36     & 53    &  \textbf{126} &  21       & 35    & 62  &  49  \\
% \textbf{\toolname}   & \textbf{3} & \textbf{13} & \textbf{17}     & \textbf{24}   & \textbf{23}      & \textbf{30}    &  138 &  \textbf{7}       & \textbf{22}    &   \textbf{46}  & \textbf{32}    \\
% % (0, 20)  & - & - & -     & -   & -      & -    &  - &  -       & -    & -  & \textbf{-}        \\
% \bottomrule
% \end{tabular}
% \label{tb:mem}
% \end{table}

\subsection{RQ4: Size and Memory Effciency}

\toolname will produce an executable program that is deployed on devices. We need to evaluate the size of the programs generated by our method to show the efficiency of \textit{Computing Unit Extraction} steps. The results are shown in Table~\ref{tb:size}. Our \toolname will produce programs that are much smaller than the original TFLite platform, especially for small DL models (\ie Lenet, Fruit, MobileNet). For large models, the model weights consume most of the size of the deployment files so \toolname files and original TFLite files do not have much difference.

For memory consumption, our \toolname generated program is theoretically much better than the original TFLite model, as the program produced by \toolname removes model parsing and data preparation. TFLite library needs to load the computation graph and use intermediate data to configure their operators. A random-access memory (RAM) consumption comparison among them is shown in Table~\ref{tb:mem}. Note that we do not use any memory optimization methods for all deployment methods. 

The customized DL program generated by our method is much more efficient in memory consumption than its original TFLite DL model in all cases.
% In addition, the well-known AI compiler TVM only has slightly better memory efficiency than our method in one model (\ie \ding{198} MiDaS model). However, in this case, the time consumption of TVM is much higher than the other methods (\ie 647ms in TV, 328 in \toolname). 
In this experiment, the program generated by our \toolname can significantly reduce memory consumption by 68.8\% and 36.0\% on x86-64 and ARM64 platforms, respectively, and does not increase the time taken. These benefits will help small memory edge devices to deploy larger and more models on-device.

\begin{figure}[h]
  \begin{center}
    \includegraphics[width=1.0\linewidth]{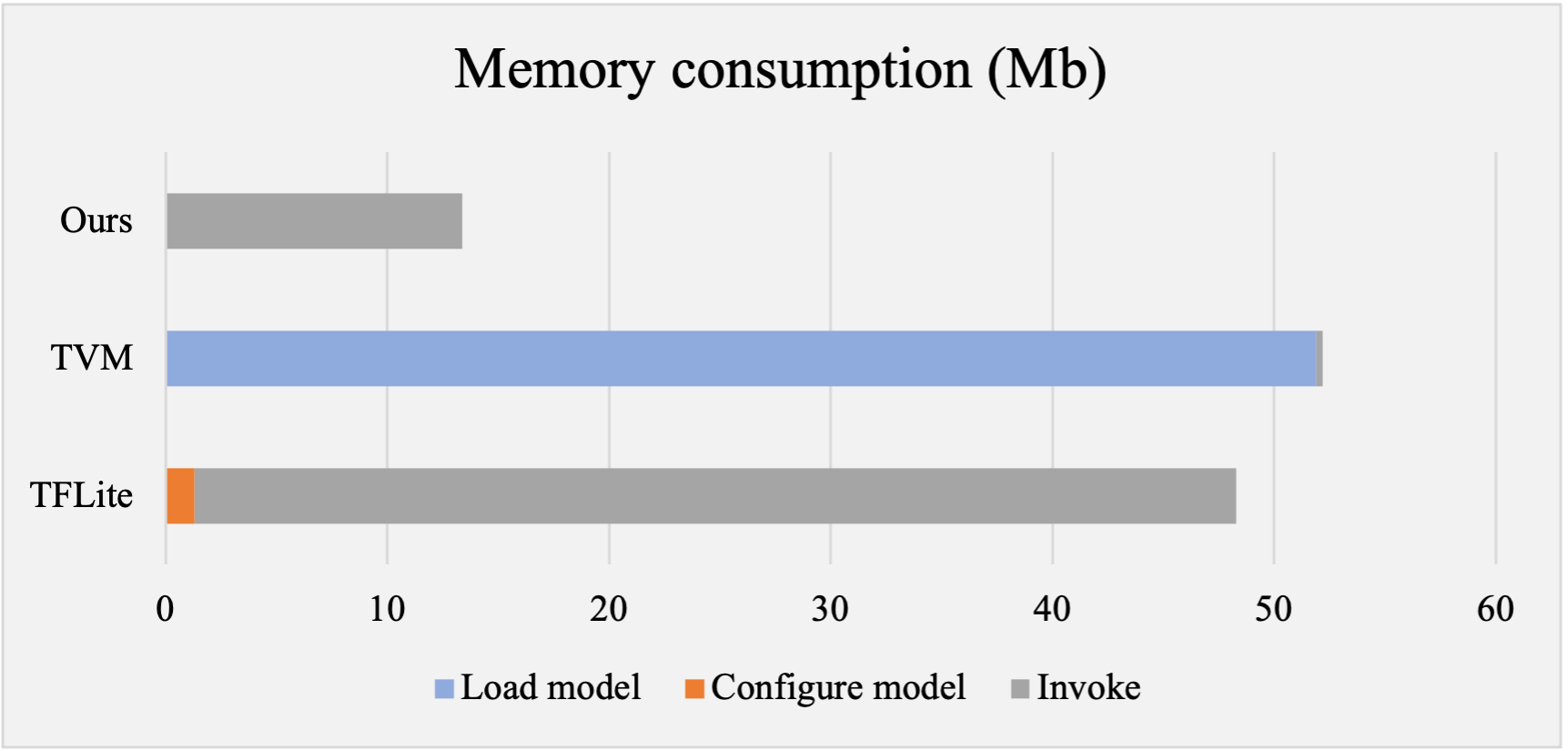}
  \end{center}
  \vspace{-1.0em}
  \caption{The pattern of memory allocation in different deployment methods on the Skin diagnosis model. A complete model inference process: load model $\rightarrow$ configure model $\rightarrow$ invoke (compute the output).}
  \label{fig:mem_inference}
\end{figure}

The three typical memory allocation patterns for different deployment methods are shown in Figure~\ref{fig:mem_inference}. TVM allocates memory mainly on the model loading step, which means it will allocate data (\eg wights) in memory before the model inference and keep them in the memory. For TFlite, it allocates some tensors when configuring the computing function. However, our method only allocates the memory in the invoking step, \ie computing the outputs, because it does not parse the model representation and only allocates tensors in the computing process. 

\begin{tcolorbox}[colback=gray!5!white,colframe=gray!85]
RQ4 Answer: The C++ program produced by \toolname uses much less memory than the original TFLite model in all cases. It is especially memory efficient on the large language model example, \ie GPT-2. 
\end{tcolorbox}

\begin{table*}[t!]
% \small
\small
% \vspace{-1.0em}
\centering
\caption{The model inference time (\ie latency), and RAM consumption (Mb) of our proposed \toolname and other model deployment strategies on \texttt{x86-64} platform. \toolname program, TVM model, and ONNX-Runtime model are all converted from the TFLite model.}
\begin{tabular}{c|lccccccccccc|c}
\hline
\multirow{4}{*}{Latency} & Model name    & Fruit & Skin cancer &MobileNet&MNASNet &SqueezeNet &EfficientNet &MiDaS &Lenet &PoseNet &SSD & GPT-2 & Average \\
\cline{2-14}
& TVM     & \textbf{28.1} & 82.4 & 61.3     &  85.0  & 69.9      & 82.2    &  447.6 &  26.6       & 103.3    & 207.7 & 468.3 & 151.4  \\
& ONNX-Runtime     & 31.6 & \textbf{64.9} & 59.8     &  76.4  & 54.1     &  \textbf{79.7}   & \textbf{285.6}  & 27.1        &  \textbf{89.5}   & \textbf{176.3}   & 984.3  & 175.4 \\
& \textbf{\toolname}  & 32.6 & 91.9 & \textbf{54.6}     & \textbf{70.8}   & \textbf{50.2}      & 84.9    &  317.0 &  \textbf{2.0}       & 110.3    &   195.9  & \textbf{433.0} & \textbf{131.2}   \\
% (0, 20)  & - & - & -     & -   & -      & -    &  - &  -       & -    & -  & \textbf{-}        \\
\hline
\multirow{3}{*}{RAM}& TVM     & 35.50 & 61.93 & 47.05     &  65.09  &40.78     & 70.31    &  179.62 &  37.85       & 43.97    & 93.61 & 1115.14 &  162.80   \\
& ONNX-Runtime     & 25.97  & 53.81 & 42.31     & 48.91   & 25.80     &  48.68   & 154.25  &  27.00       &  42.82   & 76.48   & 1248.91  & 163.18\\
& \textbf{\toolname}  & \textbf{8.89} & \textbf{11.95} & \textbf{10.47}     & \textbf{9.65}   & \textbf{17.74}      & \textbf{11.95}    &  \textbf{94.30} &  \textbf{5.96}       & \textbf{14.73}    &   \textbf{19.01} & \textbf{94.99} & \textbf{27.24}   \\
\hline
\end{tabular}
\label{tb:compare_others}
\end{table*}
% \subsection{RQ3: Overhead}
% Compared with other

% \begin{table*}[t]
% % \small
% \small
% % \vspace{-1.0em}
% \centering
% \caption{The maximal compilation error $(\times 10^{-5})$ of \toolname. \toolname and TVM programs are all converted from the TFLite model. Our method does not have any compilation errors as \toolname use the same computing process as the original model inference. }
% \begin{tabular}{lccccccccccc|c}
% \hline
%  Model name    & Fruit & Skin cancer &MobileNet&MNASNet &SqueezeNet &EfficientNet &MiDaS &Lenet &PoseNet &SSD & GPT-2 & Average \\
% \hline
% TVM     & 0.4 & 0.13 & 0.21     &  1.03  & 0.17      & 0.07    &  73.24 &  0.05       & 4.20    & 1.86   &   & \\
% ONNX-Runtime     &  &  &      &    &      &     &   &         &     &    &   & \\
% \hline
% \toolname  & 0 & 0 & 0     & 0   & 0      & 0    &  0 &  0       & 0    &   0   &   & \\
% % (0, 20)  & - & - & -     & -   & -      & -    &  - &  -       & -    & -  & \textbf{-}        \\
% \hline
% \end{tabular}
% \label{tb:correctness}
% \end{table*}

\subsection{RQ5: Comparison with Other Strategies}
%Our method can compile the TFLite model into an executable program. 
The RQ3 and RQ4 show that the generated program runs much more efficiently than the original model inference of TFLite. In this research question, we would like to compare our tool and other model deployment strategies like TVM and ONNX-Runtime to show the effectiveness of our \toolname method. Although TVM supports packing the model information into the API library, they both need the model representation to execute the model inference and not pure code implementation. We have compared the compilation error of the proposed method, TVM, and ONNX-Runtime in RQ1. 

For latency, our \toolname slightly outperforms or underperforms the other model deployment strategies. It does significantly better on the GPT-2 LLM  (see Table~\ref{tb:compare_others}, \emph{Latency} row). 

For memory, our method performs significantly better than other strategies, \ie reduces the RAM consumption by more than 80\% on \texttt{x86-64} platform (see Table~\ref{tb:compare_others}, \emph{RAM} row).  Programs generated by our \toolname approach are very memory-efficient. For example, on a large language model ML example (\ie GPT-2), it consumes 10 times less RAM than other strategies. On all others, it ranges from 1.4 times less to 6 times less.

\begin{tcolorbox}[colback=gray!5!white,colframe=gray!85]
RQ5 Answer: \toolname significantly outperforms other compilation strategies, \ie TVM and ONNX-Runtime, in terms of RAM consumption. In addition, our method achieves comparable performance in terms of latency.
\end{tcolorbox}

\section{Discussion}
In this section, we will discuss how to maintain our tool, the meta-model of our method, and its limitations.

\subsection{Integrating Our Method into Existing Tools}

\begin{figure}[t]
  \begin{center}
    \includegraphics[width=0.9\linewidth]{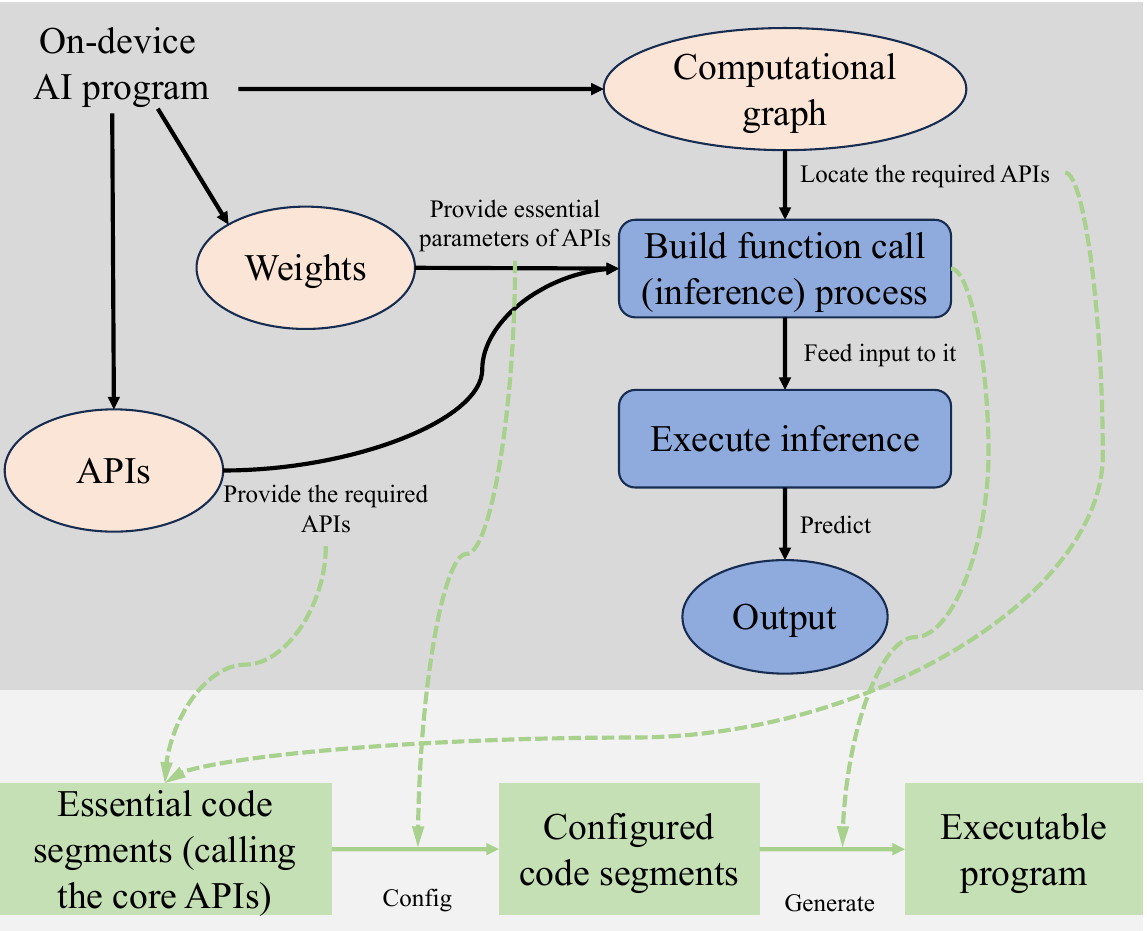}
  \end{center}
  \vspace{-0.7em}
  \caption{Meta-model our method. Model representations including computational graphs and weights can be stored as a separate file or be integrated into the API library. Because AI platforms usually are open-sourced, the source code of API libraries can be collected from the Internet.}
  \label{fig:meta}
\end{figure}

We developed our method based on current function call process and backend APIs of TFLite. We may need to update some parts of our method, \eg code extraction and configuration, to support new versions of TFLite. Our approach would be better if integrated into existing on-device DL libraries such as TFLite and other AI compilers. Our prototype tool cannot currently support all DL models. However, the on-device DL libraries have similar inference workflows. We provide a design meta-model (see Figure~\ref{fig:meta}) that developers can use to integrate our research ideas into their public tools. In existing on-device AI programs, the computational graph will be parsed to locate the required APIs in the library. Then, the model weights will be loaded into the computing function to build a complete inference process. After obtaining the complete inference process, the input data will be fed to it to get outputs. To generate pure code implementations for on-device model inference, the essential code segments (usually C/C++ or assembly codes) can be located by tracing the function call when building the function call process in AI programs. Each code segment contains the complete computing process from input to output of one specific operator in the on-device models. Next, the model weights can be extracted and parsed from the model representation, and they can be used to configure the extracted codes, \eg creating essential variables or instants for the C++ functions. The process of generating the complete program based on the configured code segments is then similar to building an inference process in AI programs, \ie call the functions that include each operator's code segment as the order of operators in the on-device model.

\subsection{Limitations}
Our \toolname is designed for on-device platforms like TFLite. Our methods may not be effective on other DL platforms without an optimized on-device implementation, as the extracted code will be inefficient if the DL library implements training-related code in operators. However, users can use model conversion tools to convert other models to TFLite models and then use our tool.

% We conducted our experiments on 11 representative ML models commonly deployed to mobile devices using TFLite. However, our results may not generalize to other ML models or other mobile deployment platforms.

We have not evaluated our \toolname with real-world mobile ML developers. 
We have not evaluated our method with side-channel information (\eg RAM, CPU usage) to reconstruct the model~\cite{wei2020leaky,duddu2018stealing,liu2019side}. Our method may be effective for them because the generated code has a different RAM usage pattern from the original model (see Figure~\ref{fig:mem_inference}). However, our generated program has the same CPU usage pattern as the original model.
        \section{Conclusion}

In this paper, we analyze the key security issues in existing model deployment strategies. Attackers can identify the DL components (\eg models, libraries, programs) on devices, and generate effective attacks against these DL models. We propose a method~\toolname to extract the computing codes of the DL model, refactor the extracted codes, and compile them into an executable program. Our proposed \toolname has four steps including \firstmodule, \secondmodule, \thirdmodule, and \forthmodule. Our \toolname will generate complete custom programs of model inference without the need for model representations. Our experiments show that our method not only achieves a higher level of security compared with existing methods but also can accelerate the model inference and reduce memory usage. In future, we will analyze the source code of the DL library to optimize extracted code to improve efficiency of the executable program.

\section{Data-Availability Statement}

We provide a GitHub repository of our artifact~\cite{Zhou2024customdlcoder}. It has the instruction of installing the dependency and testing our tool.

\begin{acks}
This work is partially supported by the Open Foundation of Yunnan Key Laboratory of Software Engineering under Grant No.2023SE102, by the National Natural Science Foundation of China under Grant No.62202026 and No.62172214, and by ARC Laureate Fellowship FL190100035. 
\end{acks}
        
	\bibliographystyle{ACM-Reference-Format}
	\bibliography{acmart}

\end{document}